\documentclass[11pt]{cernrep}

\usepackage{graphicx}

\usepackage{mcite}

\begin{document}
\title{From HERA to the LHC}
\author{John Ellis}
\institute{CERN, Geneva}
\maketitle

\begin{flushright}
CERN-PH-TH/2005-243
\end{flushright}

\begin{abstract}
Some personal comments are given on some of the exciting interfaces
between the physics of HERA and the LHC. These include the quantitative understanding
of perturbative QCD, the possible emergence of saturation phenomena and the Colour-Glass
Condensate at small $x$ and large $Q^2$, the link between forward physics and
ultra-high-energy cosmic rays, and new LHC opportunities opened up by the discovery
of rapidity-gap events at HERA, including the search for new physics such as Higgs bosons
in double-diffraction events.
\end{abstract}

\section{Preview}

There are many exciting interfaces between physics at HERA and the LHC,
and I cannot do justice to all of them in this talk. Therefore, in this
talk I focus on a few specific subjects that interest me personally,
starting with the LHC's `core business', namely the search for new physics
at the TeV scale, notably the Higgs boson(s) and
supersymmetry~\cite{newLHC}.  Identifying any signals for such new physics
will require understanding of the Standard Model backgrounds, and QCD in
particular. I then continue by discussing some other topics of specific interest
to the DESY community.

$\bullet$ The understanding of QCD will be important for making accurate studies of any such new
physics. Perturbative QCD at moderate $x$ and large $p_T$ is quite well understood, with dramatic
further progress now being promised by novel calculational techniques based on string 
theory~\cite{stringQCD}.

$\bullet$ Novel experimental phenomena are now emerging at RHIC at small $x$, following 
harbingers at HERA. The parton density saturates, and a powerful organizational framework 
is provided by the Colour-Glass Condensate (CGC)~\cite{CGC}. 
Forward measurements at the LHC will provide unique opportunities for following
up on this HERA/RHIC physics.

$\bullet$ Forward physics at the LHC will also provide valuable insight into the interpretation of
ultra-high-energy cosmic rays (UHECRs)~\cite{UHECRs}. 
One of the principal uncertainties in determining their
energy scale is the modeling of the hadronic showers they induce, and the LHC will be the closest
laboratory approximation to UHECR energies.

$\bullet$ Looking further forward, there is increasing interest in
exploring at the LHC the new vistas in hard and soft diffraction opened up
by the discovery of rapidity-gap events at HERA~\cite{HERAgap}. One
particularly interesting possibility is quasi-exclusive diffractive
production of Higgs bosons or other new particles at the
LHC~\cite{Durham}.  This is particularly interesting in supersymmetric
extensions of the Standard Model, notably those in which CP is
violated~\cite{ELP}.

\section{Prospects in Higgs Physics}

Many studies have given confidence that the Standard Model Higgs boson
will be found at the LHC, if it exists~\cite{LHCHiggs}.  Moreover, there
are some chances that it might be found quite quickly, in particular if
its mass is between about 160~GeV and 600~GeV. However, discovering the
Higgs boson will take rather longer if its mass is below about 130~GeV, as
suggested in the minimal supersymmetric extension of the Standard Model
(MSSM)~\cite{MSSMHiggs}.  In this case, the Higgs signal would be composed
of contributions from several different production and decay channels,
notably including $gg \to H \to \gamma \gamma$.

Understanding the gluon distribution at $x \sim 10^{-2}$ is therefore a high priority, and one to
which HERA measurements of processes involving gluons have been playing key roles~\cite{PDF}.
Perturbative corrections to the $gg \to H$ production process need to be
understood theoretically, as do the corrections to $H \to \gamma \gamma$ decay. Resummation of the
next-to-next-to-leading logarithms has by now reduced these uncertainties to the 10\% 
level, and further improvements may be possible with the string-inspired
calculational techniques now being introduced~\cite{HMHV}.

\begin{figure}[htb!]
\begin{center}  
\includegraphics[width=.48\textwidth]{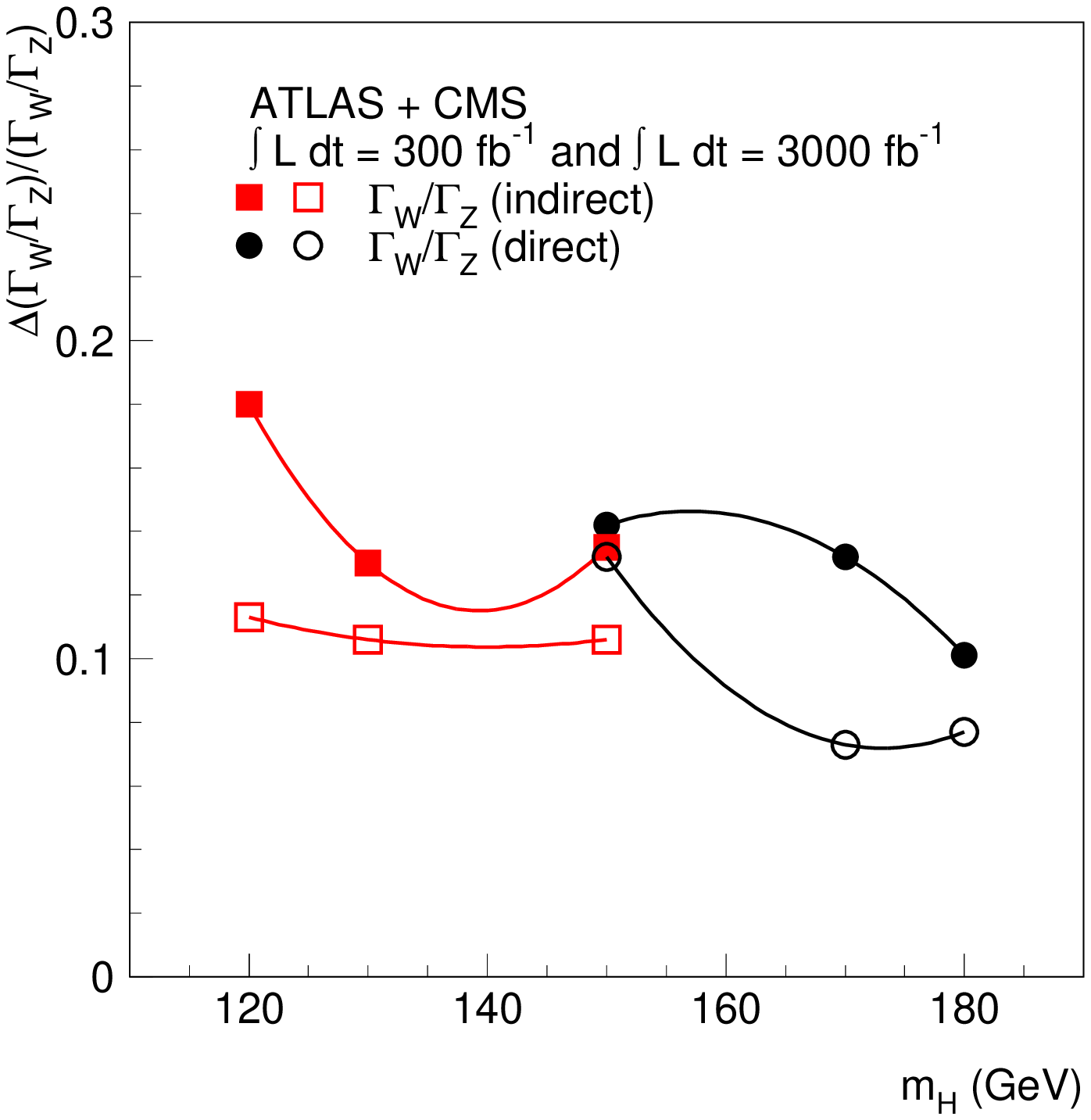}
\includegraphics[width=.48\textwidth]{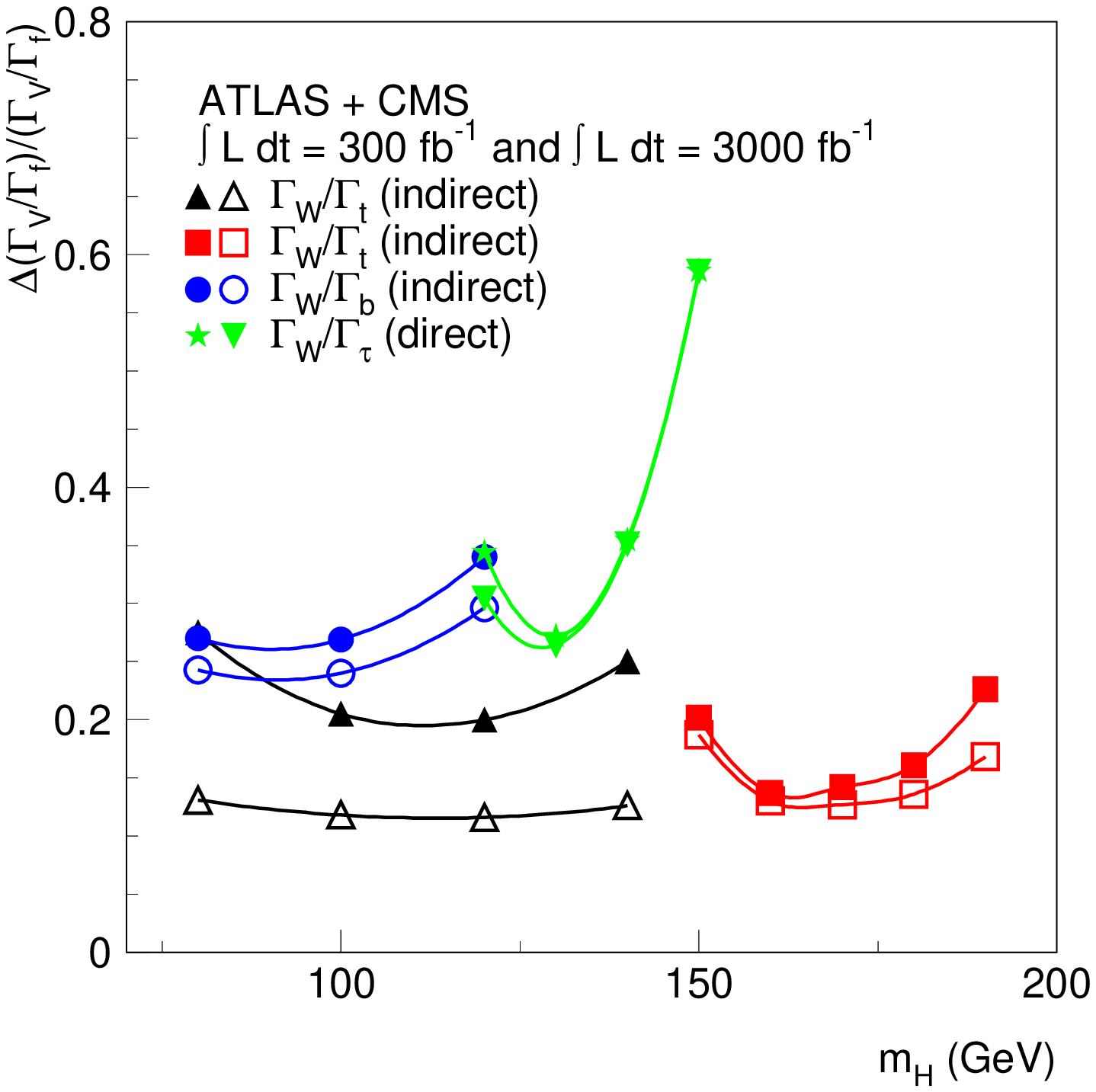}
\end{center}
\caption{
Illustrations of the accuracy with which Higgs couplings could be measured 
at the LHC with the planned luminosity and with a possible upgrade by a 
factor of ten (SLHC)~\protect\cite{Higgscouplings}.  
}
\label{fig:accuracy}
\end{figure}

Fig.~\ref{fig:accuracy} shows estimates of the accuracy with which various
Higgs couplings may be determined at the LHC, also if the luminosity may be
increased by an order of magnitude (SLHC)~\cite{Higgscouplings} [see also~\cite{LHCILC}].  
There are interesting prospects for measuring the couplings to $\tau \tau,
{\bar b} b, WW, ZZ$ and ${\bar t} t$ as well as the total Higgs decay
width, though not with great accuracy. Measurements at the ILC would
clearly be much more powerful for this purpose~\cite{LHCILC}.

\section{Theorists are Hedging their Bets}

The prospect of imminent Higgs discovery is leading theorists to place their last bets on the LHC
roulette wheel, and many are hedging their bets by proposing and discussing alternatives to the
Standard Model or the MSSM. Composite Higgs models are not greatly favoured, since they have a
strong tendency to conflict with the precision electroweak data~\cite{LEPEWWG}. 
This problem has led some
theorists to question the interpretation of the electroweak data, which are normally taken to
favour $m_H < 300$~GeV, debating their consistency and even arguing that some data should
perhaps be discounted~\cite{Chanowitz}. Personally, I see no strong reason to doubt the hints
from the electroweak data. An alternative corridor leading towards higher Higgs masses is provided by
including higher-dimensional operators in the electroweak data analysis~\cite{corridor}: this would
require some fine-tuning, but cannot be excluded. An even more extreme
alternative that has been re-explored recently is that of Higgsless models~\cite{Hless}. 
However, these lead to
strong $WW$ scattering and conflict with the available electroweak data. These problems are
alleviated, but not solved, by postulating extra dimensions at the TeV scale~\cite{extraDHless}.

One of the least unappetizing alternatives to the supersymmetric Higgs paradigm is offered by
little Higgs models~\cite{lHiggs}. 
Their key idea is to embed the Standard Model in a larger gauge group, from
which the Higgs boson emerges as a relatively light pseudo-Goldstone boson. The one-loop quadratic
divergence due to the top quark:

$$
\delta m^2_{H,top} (SM) \sim (115 \; {\rm GeV})^2  \,
\left ( \frac{\Lambda}{400 \; {\rm GeV}} \right )^2
$$
is cancelled by the contribution of a new heavy $T$ quark:

$$
\delta m^2_{H,top} (LH) \sim
\frac{6 G_F m^2_t}{\sqrt{2} \pi^2} \, m^2_T \,
\log \frac{\Lambda}{m_T}
$$
Additionally, there are new gauge bosons and exotic Higgs representations. The Standard-Model-like
Higgs boson is expected to be relatively light, possibly below $\sim 150$~GeV, whereas the other
new particles are expected to be heavier:

$$
\begin{array}{lll}
M_T & < & 2 \, {\rm TeV} (m_h / 200 {\rm GeV})^2 \\
M_W' & < & 6 \, {\rm TeV} (m_h / 200 {\rm GeV})^2 \\
M_{H^{++}} & < & 10 \, {\rm TeV}
\end{array}
$$
Certainly the new $T$ quark, probably the $W'$ boson and possibly even the
doubly-charged Higgs boson will be accessible to the LHC.  Thus little
Higgs models have quite rich phenomenology, as well being decently
motivated. However, they are not as complete as supersymmetry, and would
require more new physics at energies $> 10$~TeV.
 
Depending on the mass scale of this new physics, there may be some
possibility for distinguishing a little Higgs model from the Standard
Model by measurements of the $gg \to H \to \gamma \gamma$ process at the
LHC.  However, the ILC would clearly have better prospects in this 
regard~\cite{LHCILC}.

\section{Supersymmetry}

No apologies for repeating the supersymmetric mantra: it resolves the
naturalness aspect of the hierarchy problem by cancelling systematically
the quadratic divergences in all loop corrections to the Higgs mass and
hence stabilizes the electroweak scale~\cite{hierarchy}, it enables the
gauge couplings to unify~\cite{GUT}, it predicts $m_H <
150$~GeV~\cite{MSSMHiggs} as suggested by the precision electroweak
data~\cite{LEPEWWG}, it stabilizes the Higgs potential for low Higgs
masses~\cite{ER}, and it provides a plausible candidate~\cite{EHNOS} for the dark matter that
astrophysicists and cosmologists claim clutters up the
Universe.

However, all we have from accelerators at the moment are lower limits on the possible supersymmetric particle
masses, most notably from the absence of sparticles at LEP: $m_{\tilde \ell}, m_{\chi^\pm} > 100$~GeV and the
Tevatron collider: $m_{\tilde g}, m_{\tilde q} > 300$~GeV, the LEP lower limit $m_H > 114.4$~GeV, and the
consistency of $b \to s \gamma$ decay with the Standard Model. However, if we assume that the astrophysical cold
dark matter is largely composed of the lightest supersymmetric particle (LSP), and require its density to lie
within the range allowed by WMAP et al~\cite{WMAP}:

$$
		0.094 < \Omega_{\chi} h^2 < 0.129,
$$
we obtain upper as well as lower limits on the possible sparticle masses. The anomalous magnetic
moment of the muon, $g_\mu - 2$, provides intermittent hints on the supersymmetric mass scale~\cite{g-2}: 
these are lower
limits if you do not believe there is any significant discrepancy with the Standard Model prediction, but also an
upper limit if you do not believe that the Standard Model can fit the data, as is indicated by the current
interpretation of the $e^+ e^-$ data used to calculate the Standard Model prediction.

\begin{figure}[htb!]
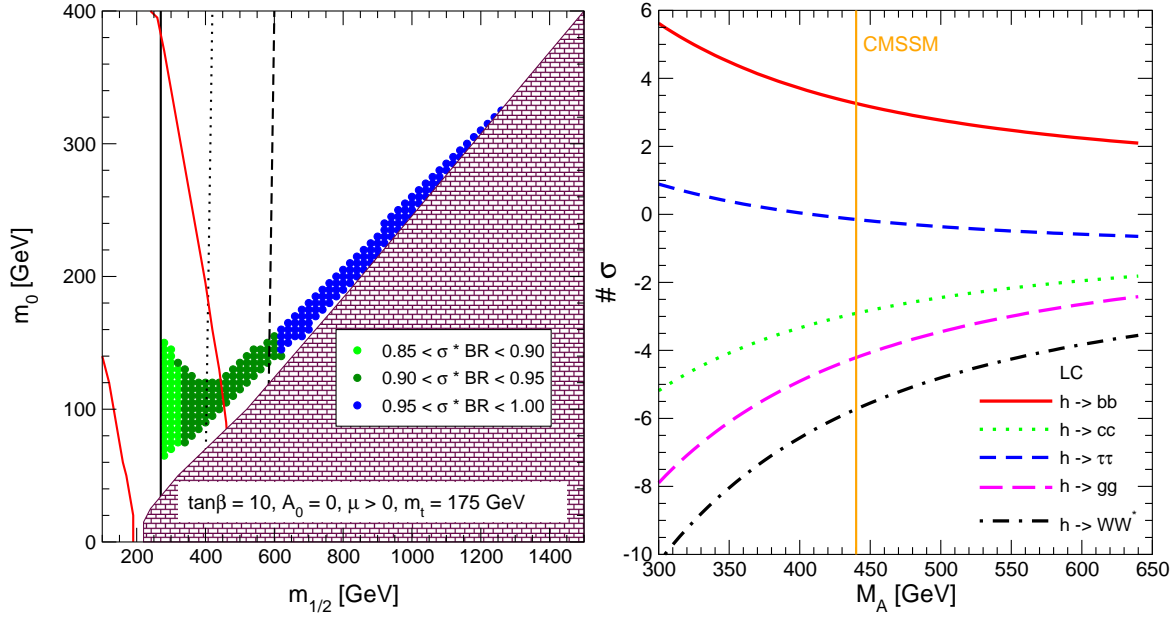

\begin{center}  
\includegraphics[width=.48\textwidth]{EHOW03c.03.cl.eps}
\includegraphics[width=.48\textwidth]{EHOW2.NUHM_MA01.cl.eps}
\end{center}
\caption{
Left panel: The cross section for production of the lightest CP-even MSSM
Higgs boson in gluon fusion and its decay into a photon pair, $\sigma(gg
\to h) \times {\cal B}(h \to \gamma\gamma)$, normalized to the Standard
Model value with the same Higgs mass, is given in the $(m_{1/2}, m_0)$
plane for $\mu > 0$, $\tan\beta = 10$, assuming $A_0 = 0$ and $m_t =
175$~GeV~\protect\cite{EHOW1}. The diagonal (red) solid lines are the $\pm 
2 -
\sigma$ contours for $g_\mu - 2$. The near-vertical solid, dotted and
dashed (black) lines are the $m_h = 113, 115, 117$~GeV contours.
The (brown) bricked regions are excluded since in these regions the LSP is the
charged $\tilde\tau_1$.
Right panel: The numbers of standard deviations by which the predictions 
of the MSSM with non-universal Higgs masses may be distinguished from 
those of the Standard Model in different channels by measurements at the 
ILC~\protect\cite{EHOW2}. The 
predictions with the CMSSM values of $M_A$ and $\mu$ are indicated by 
light vertical (orange) lines. The other
parameters have been chosen as $m_{1/2} = 300$ GeV, $m_0 = 100$ GeV,
$\tan \beta = 10$ and $A_0 = 0$.
}
\label{fig:LHChH}
\end{figure}

If one compares the production of the lightest neutral Higgs boson in the
constrained MSSM (CMSSM) in which all the soft supersymmetry-breaking
scalar masses $m_0$ and gaugino masses $m_{1/2}$ are assumed to be
universal, the {\it good news} is that the rate for $gg \to h \to \gamma
\gamma$ is expected to be within 10\% of the Standard Model value, as seen
in Fig.~\ref{fig:LHChH}(a)~\cite{EHOW1}. On the other hand, the bad news
is the rates are so similar that it will be difficult to distinguish a
CMSSM Higgs boson from its Standard Model counterpart. This would be much
easier at the ILC, as seen in Fig.~\ref{fig:LHChH}(b)~\cite{EHOW2}.

\begin{figure}[htb!]
\begin{center}
\includegraphics[width=.96\textwidth]{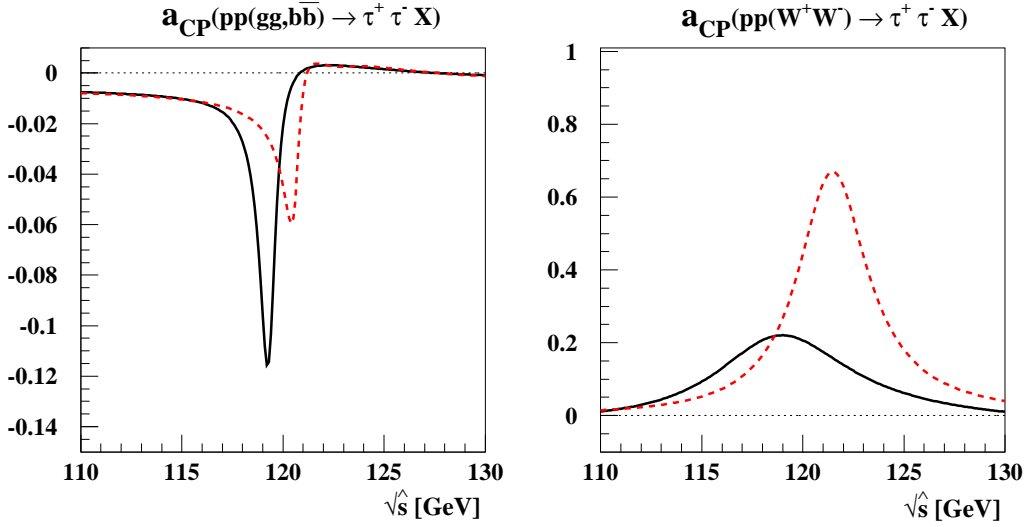}
\end{center}
\vspace{-8cm}
\caption{
Numerical  estimates   of differential   CP   asymmetries
as functions of the effective reduced centre-of-mass energy $\sqrt{{\hat
s}}$ in a CP-violating three-Higgs mixing scenario   
with gaugino phase $\Phi_3=-90^\circ$ (solid lines) and 
$\Phi_3=-10^\circ$ (dashed lines)~\protect\cite{ELP2}.
}
\label{fig:ELP2}
\end{figure}

One of the distinctive possibilities opened up by the MSSM is the
possibility of CP violation in the Higgs sector, induced radiatively by
phases in the gaugino masses and the soft supersymmetry-breaking trilinear
couplings. Fig.~\ref{fig:ELP2} displays CP-violating asymmetries that
might be observable in the $gg, {\bar b} b \to \tau^+ \tau^-$ and $W^+ W^-
\to \tau^+ \tau^-$ processes at the LHC, in one particular CP-violating
scenario with large three-way mixing between all three of the neutral MSSM
Higgs bosons~\cite{ELP2}.

\begin{figure}[htb!]
\begin{center}
\includegraphics[width=.45\textwidth]{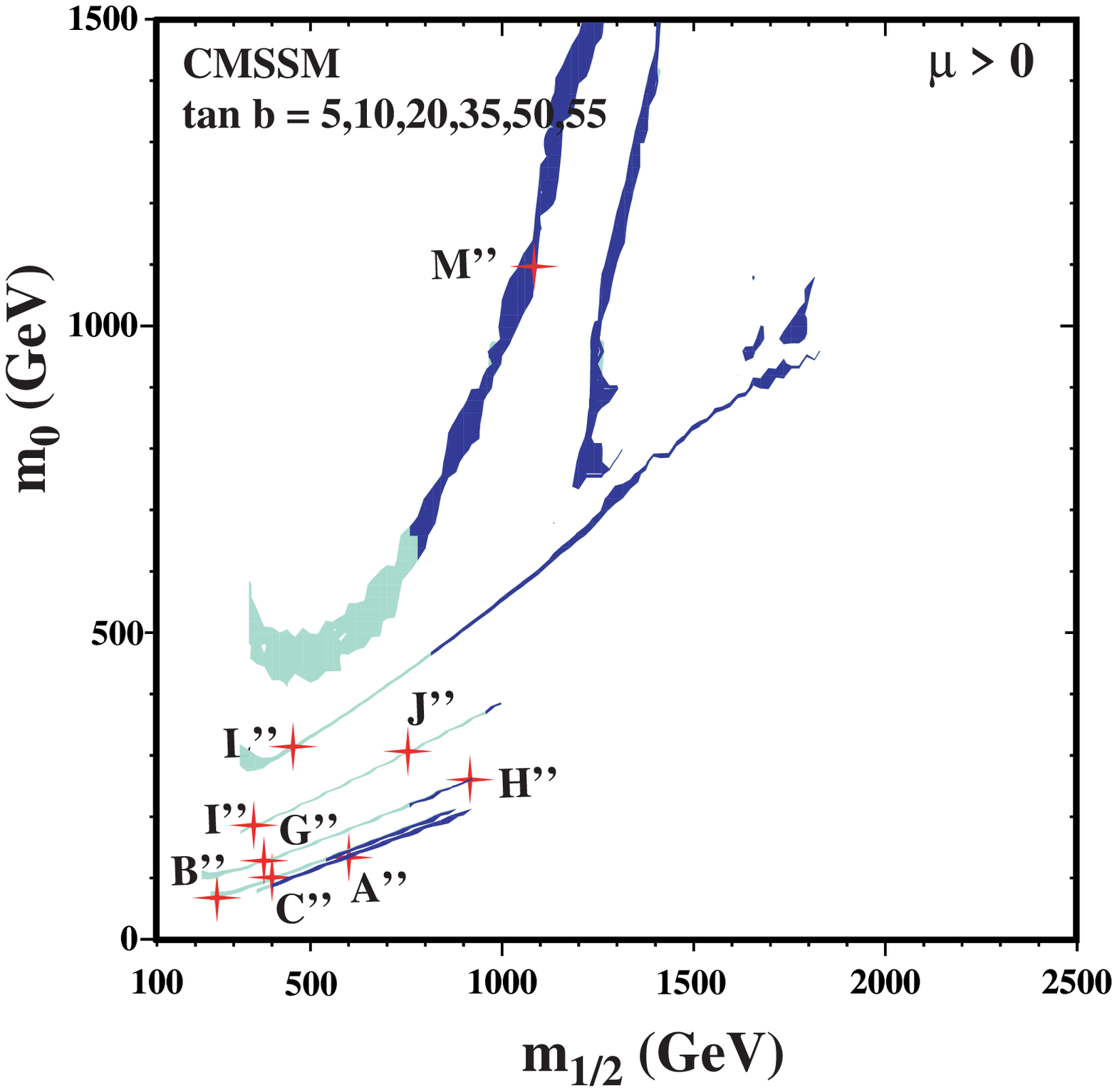}
\includegraphics[width=.45\textwidth]{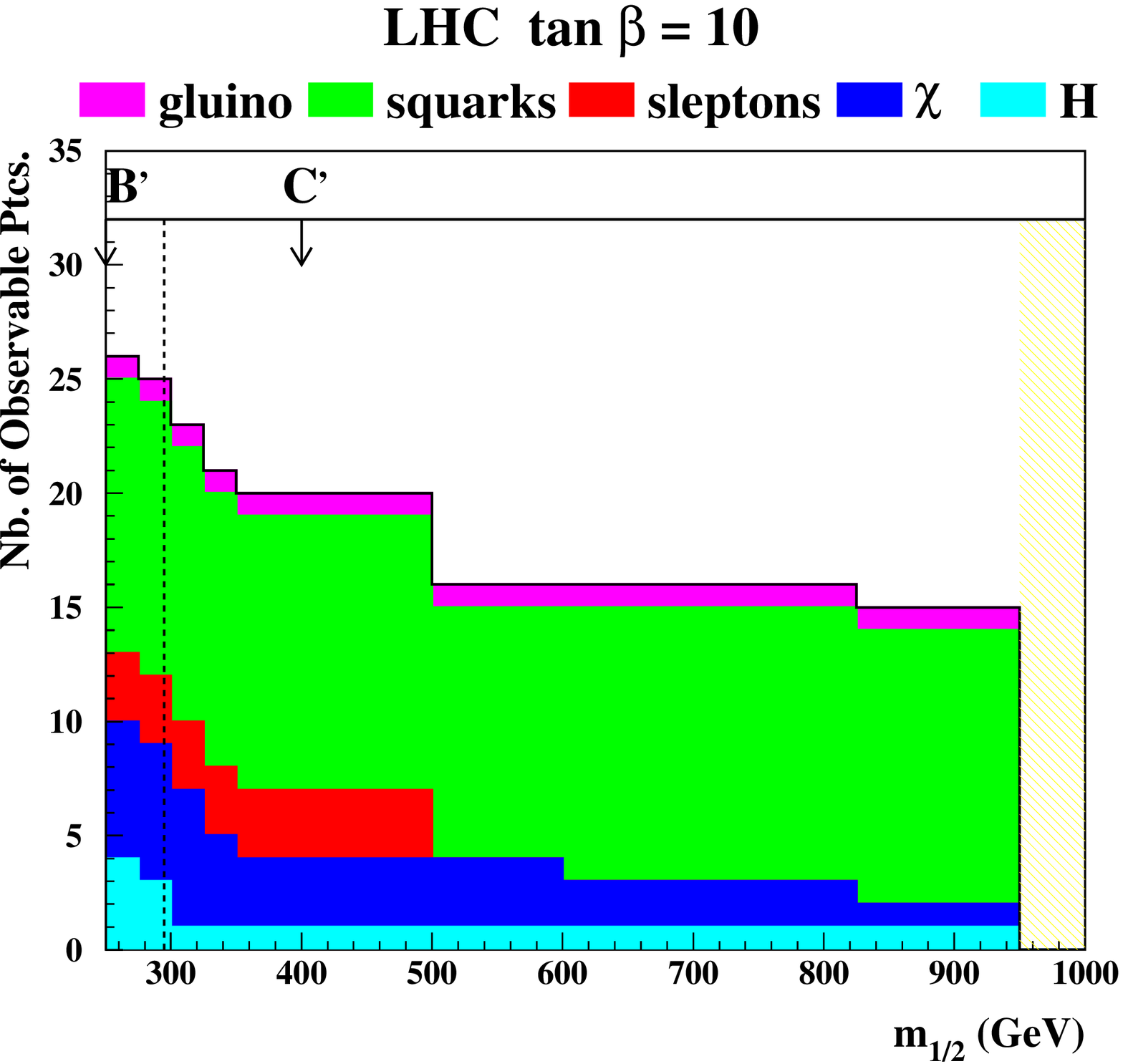}\\
\includegraphics[width=.45\textwidth]{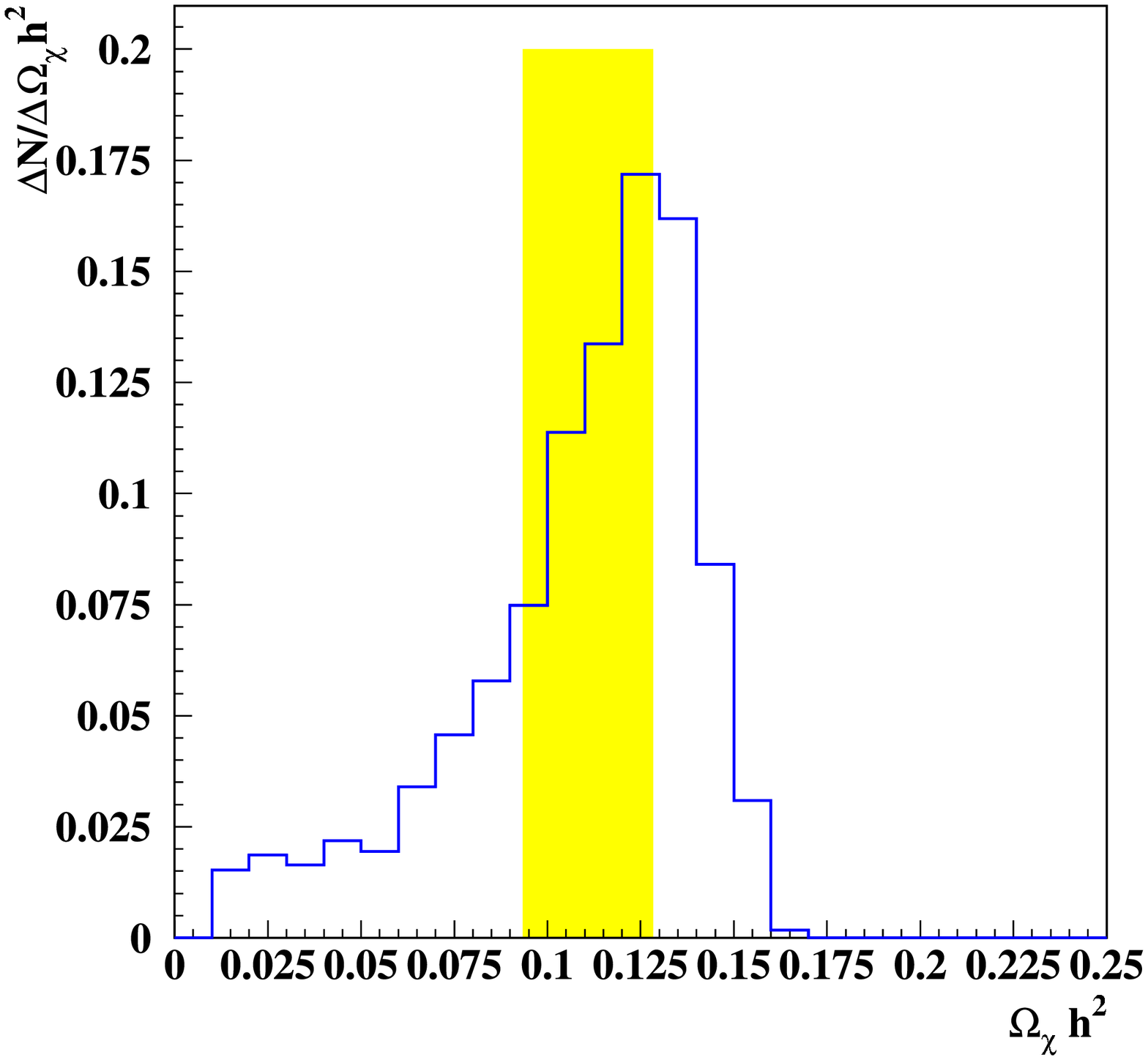}
\includegraphics[width=.45\textwidth]{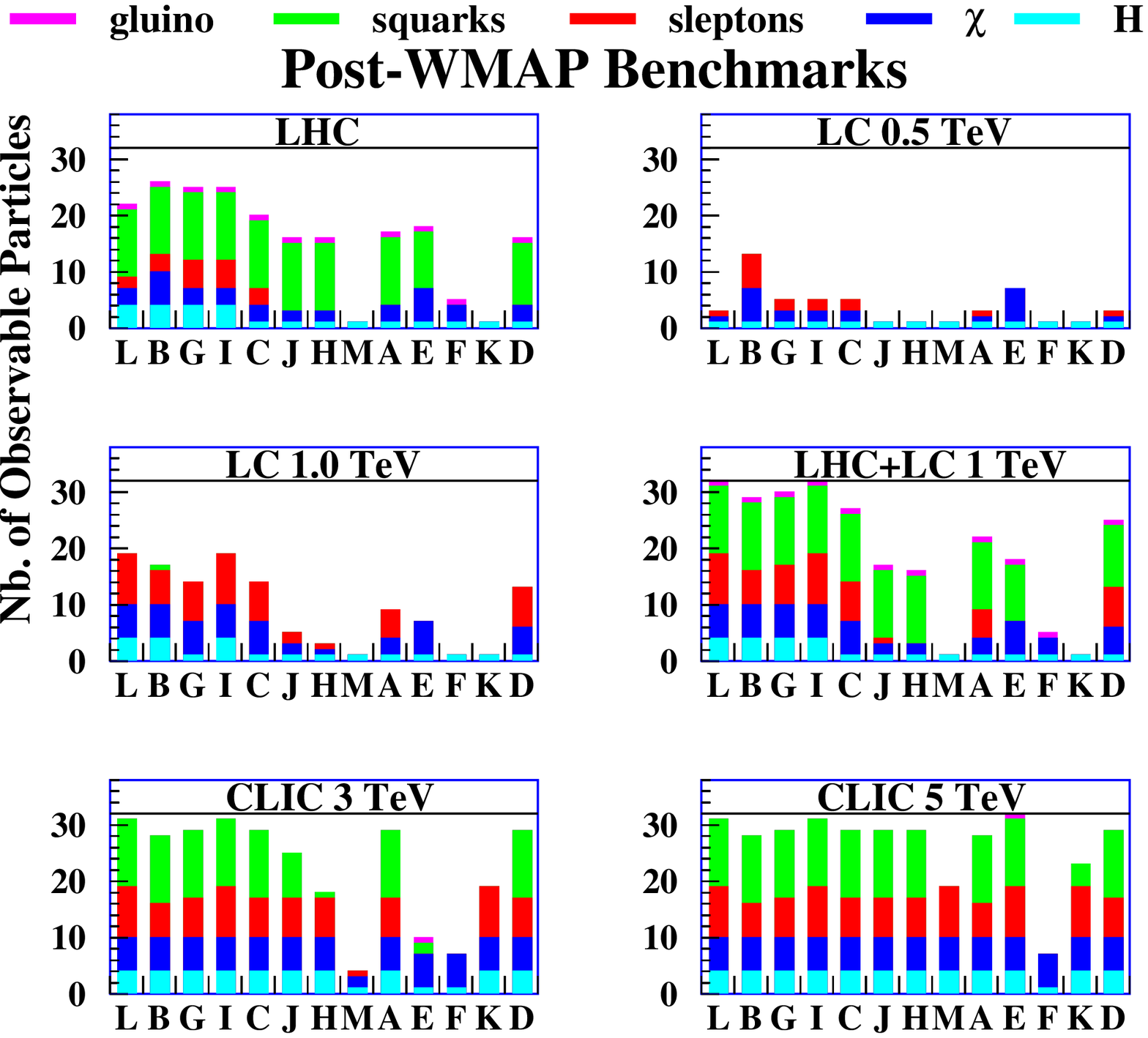}
\caption{
Top left panel: The strips of CMSSM parameter space allowed by WMAP and 
other constraints, with specific benchmark scenarios indicated by (red) 
crosses. Top right panel: The numbers of MSSM particle species observable 
at the LHC as a function of $m_{1/2}$ along the WMAP strip for 
$\tan \beta = 10$~\protect\cite{EHOW1}. Bottom left panel: The accuracy 
with which the 
relic dark matter density could be calculated using LHC measurements 
at benchmark point B, compared with the uncertainty provided by WMAP 
and other astrophysical data. Bottom right panel: The numbers of MSSM 
particle species observable in the benchmark scenarios at the LHC and 
$e^+ e^-$ colliders with different centre-of-mass 
energies~\protect\cite{EHOW2}.
}
\label{fig:Bench}
\end{center}
\vspace{-3em}
\end{figure}

A typical supersymmetric event at the LHC is expected to contain
high-$p_T$ jets and leptons, as well as considerable missing transverse
energy. Studies show that the LHC should be able to observe squarks and
gluinos weighing up to about $2.5$~TeV~\cite{LHCHiggs}, covering most of the possibilities for
astrophysical dark matter. As seen in
Fig.~\ref{fig:Bench}(a)~\cite{newLHC}, the dark matter constraint
restricts $m_{1/2}$ and $m_0$ to narrow strips extending to an upper limit
$m_{1/2} \sim 1$~TeV. As seen in Fig.~\ref{fig:Bench}(b), whatever the
value of $m_{1/2}$ along one of these strips, the LHC should be able to
observe several distinct species of sparticle~\cite{newLHC}. In a
favourable case, such as the benchmark point B in Fig.~\ref{fig:Bench}(a)
(also known as SPS Point 1a), experiments at the LHC should be able to
measure the CMSSM parameters with sufficient accuracy to calculate the
supersymmetric relic density $\Omega_\chi h^2$ (blue histogram) with
errors comparable to the present astrophysical error (yellow band) as seen
in Fig.~\ref{fig:Bench}(c)~\cite{newLHC}.  Fig.~\ref{fig:Bench}(d)
summarizes the scapabilities of the LHC and other accelerators to detect
various numbers of sparticle species. We see that the LHC is almost
guaranteed to discover supersymmetry if it is relevant to the naturalness
of the mass hierarchy. However, there are some variants of the CMSSM, in
particular at the tips of the WMAP strips for large $\tan \beta$, that
might escape detection at the LHC.

As we also see in Fig.~\ref{fig:Bench}(d), linear colliders would be able
to observe a complementary subset of sparticles, particularly sleptons,
charginos and neutralinos~\cite{newLHC}.  A linear collider with a
centre-of-mass energy of 1~TeV would have comparable physics reach to the
LHC, but a higher centre-of-mass energy, such as the 3~TeV option offered
by CLIC~\cite{CLIC}, would be needed to complete the detection and
accurate measurement of all the sparticles in most variants of the CMSSM.

\begin{figure}[htb!]
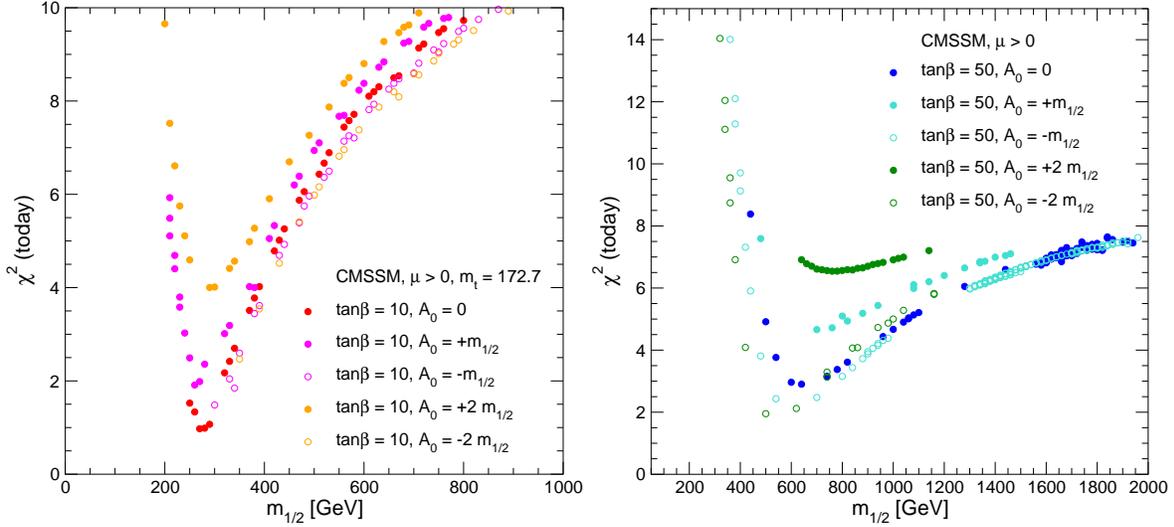

\begin{center}  
\includegraphics[width=.48\textwidth]{ehow.CHI11a.1727.cl.eps}
\includegraphics[width=.48\textwidth]{ehow.CHI11b.cl.eps}
\end{center}
\caption{
The results of $\chi^2$ fits based on the current experimental
results for the precision observables $M_W$, $\sin^2 \theta_{weff}$, 
$(g-2)_\mu$ and $b \to s \gamma$ are shown as functions of $m_{1/2}$ 
in the CMSSM parameter space with WMAP constraints for different values of
$A_0$ and (left) $m_t = 172.7 \pm 2.9$ GeV and $\tan \beta = 10$ and
(right) $m_t = 178.0 \pm 4.3$ GeV and $\tan \beta = 
50$~\protect\cite{EHOW3}.
}
\label{fig:EHOW3.3}
\end{figure}

We have recently evaluated whether precision low-energy observables
currently offer any hint about the mass scale of supersymmetric particles,
by exploring their sensitivities to $m_{1/2}$ along WMAP lines for
different values of the trilinear supersymmetry-breaking parameter $A_0$
and the ratio of Higgs v.e.v's, $\tan \beta$~\cite{EHOW3}.  The
measurements of $m_W$ and $\sin^2 \theta_W$ each currently favour $m_{1/2}
\sim 300$~GeV for $\tan \beta = 10$ and $m_{1/2} \sim 600$~GeV for $\tan
\beta = 50$. The agreement of $b \to s \gamma$ decay with the Standard
Model is compatible with a low value of $m_{1/2}$ for $\tan \beta = 10$ but prefers a
larger value for $\tan \beta = 50$, whereas $B_s \to \mu^+ \mu^-$ decay
currently offers no useful information on the scale of supersymmetry
breaking~\cite{Bmumu}. The current disagreement of the measured value of the anomalous
magnetic moment of the muon, $g_\mu - 2$, also favours independently
$m_{1/2} \sim 300$~GeV for $\tan \beta = 10$ and $m_{1/2} \sim 600$~GeV
for $\tan \beta = 50$. Putting all these indications together, as seen in
Fig.~\ref{fig:EHOW3.3}, we see a preference for $m_{1/2} \sim
300$~GeV when $\tan \beta = 10$, and a weaker preference for $m_{1/2} \sim
600$~GeV when $\tan \beta = 50$~\cite{EHOW3}.  At the moment, this
preference is far from definitive, and $m_{1/2} \to \infty$ is excluded at
lass than 3 $\sigma$, but it nevertheless offers some hope that
supersymmetry might lurk not far away.

\begin{figure}[htb!]
\begin{center}
\includegraphics[width=.45\textwidth,height=5.4cm]{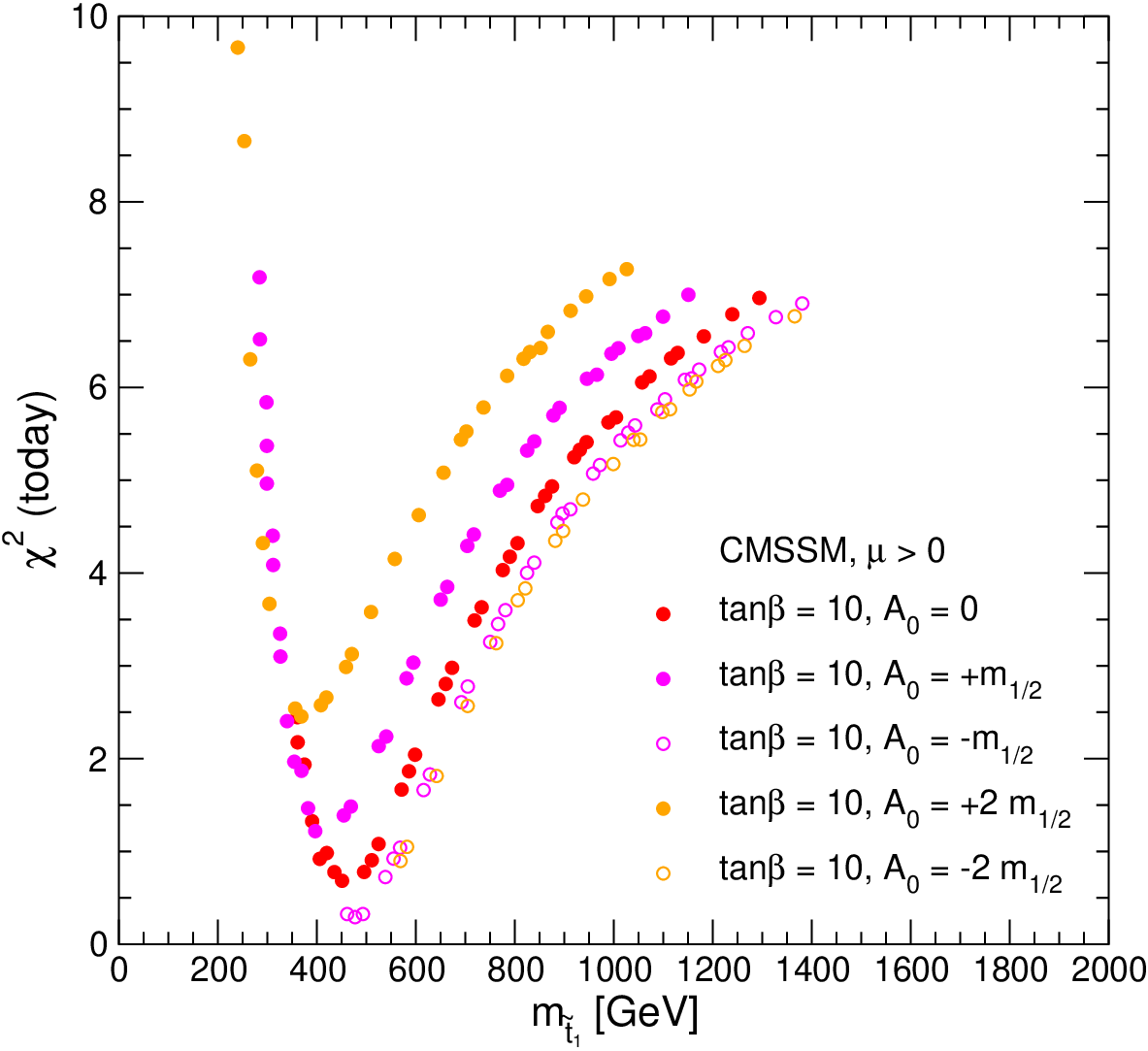}
\includegraphics[width=.45\textwidth,height=5.4cm]{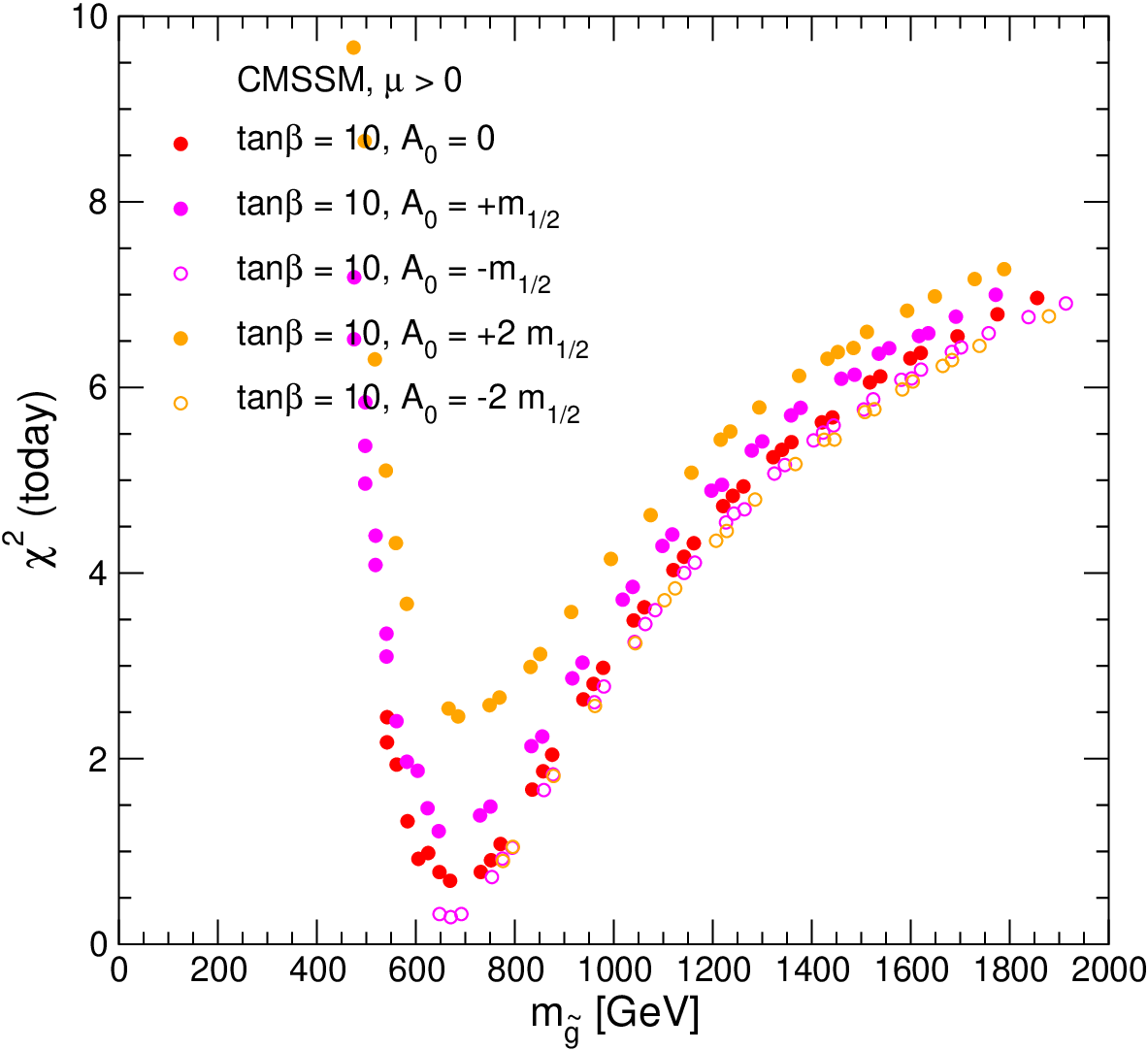}\\
\includegraphics[width=.45\textwidth,height=5.4cm]{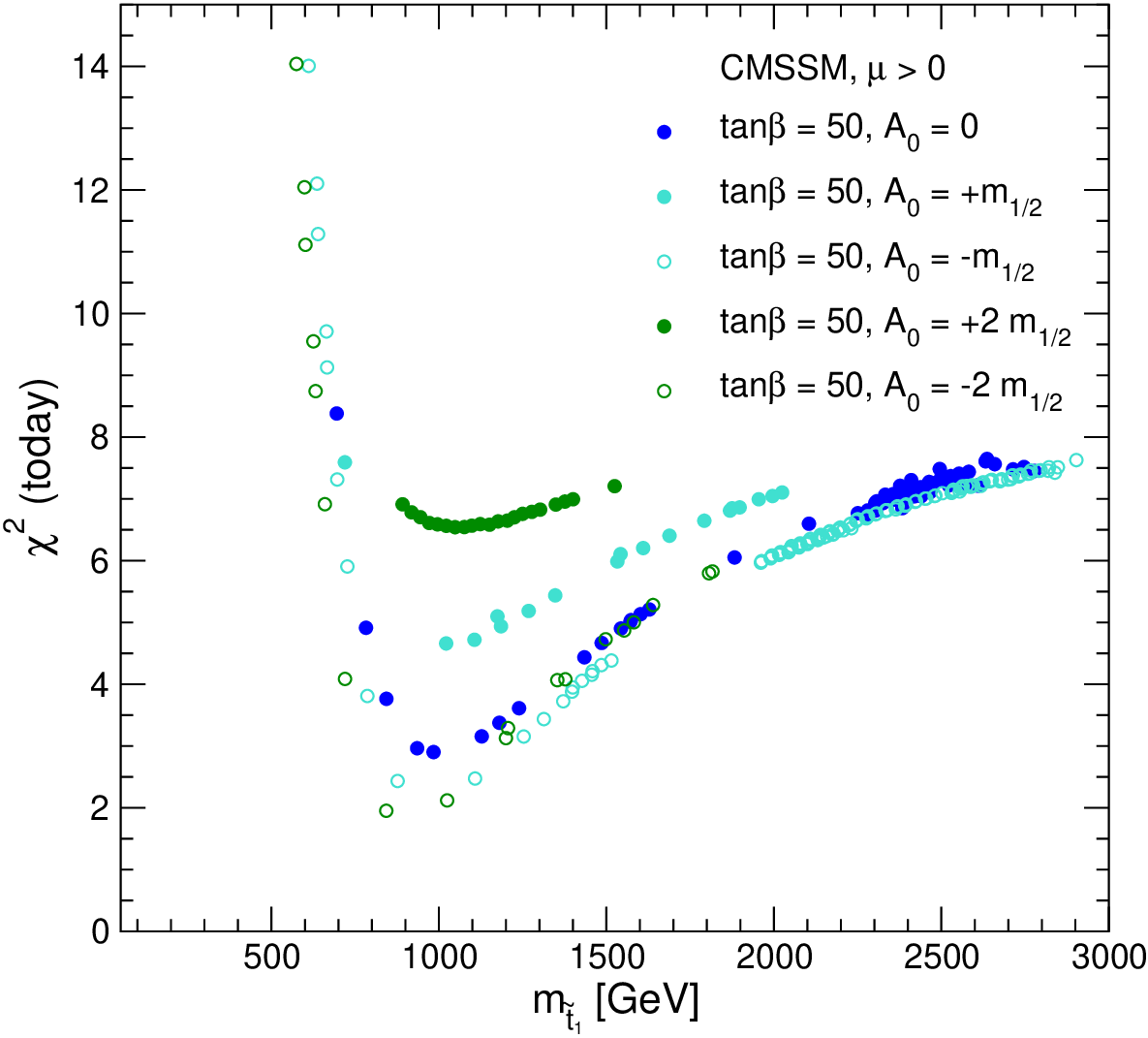}
\includegraphics[width=.45\textwidth,height=5.4cm]{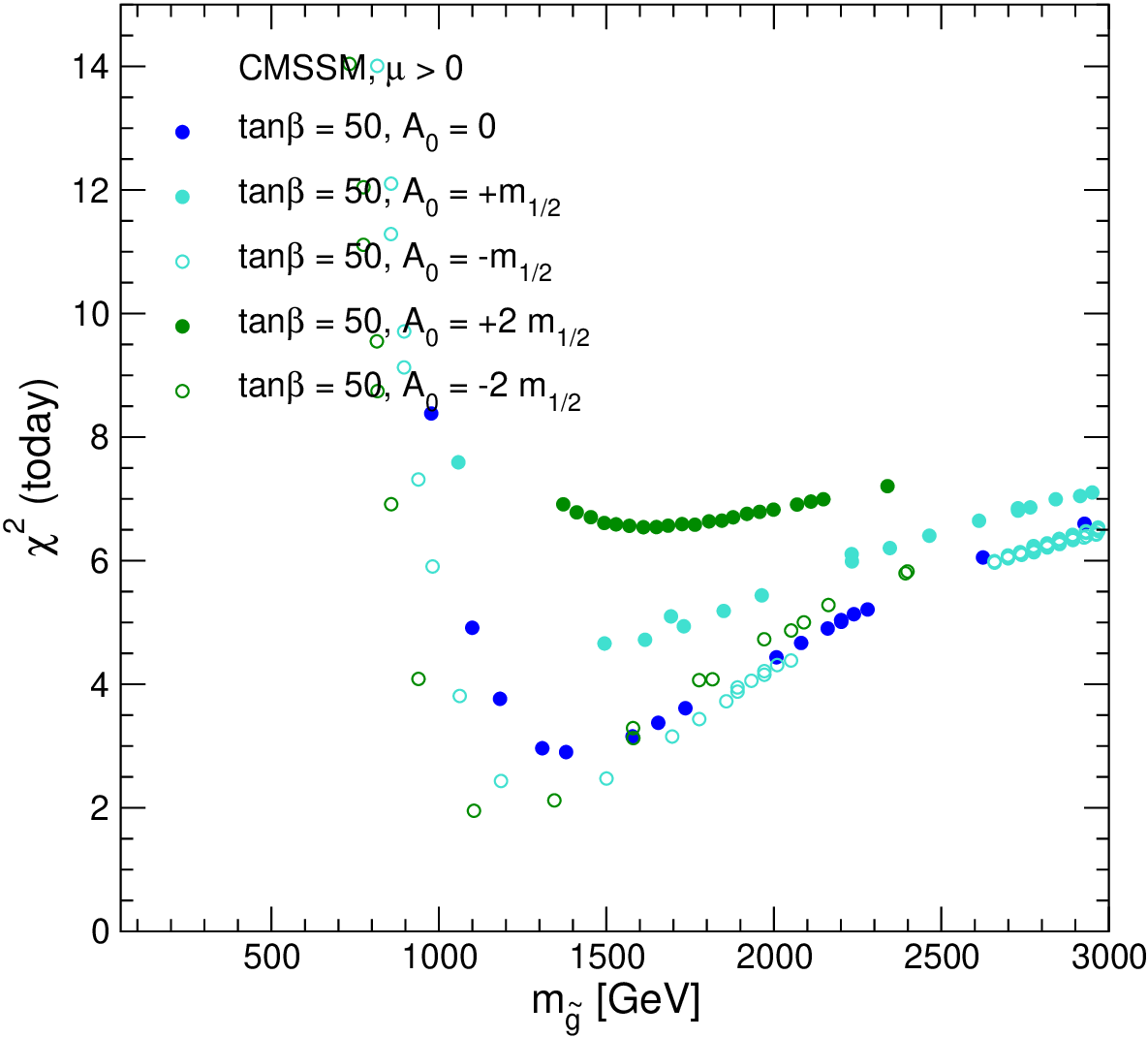}
\caption{
The $\chi^2$ contours in the CMSSM with $\tan \beta = 10$ for the lighter stop 
(left) and gluino (right) masses, assuming $\tan \beta = 10$ (top) and 
$\tan \beta = 50$ (bottom)~\protect\cite{EHOW3}.
} 
\label{fig:EHOW3.4}
\end{center}
\vspace{-3em}
\end{figure}

As seen in Fig.~\ref{fig:EHOW3.4}, the likelihood function for $m_{1/2}$
can be converted into the corresponding likelihood functions for the
masses of various species of sparticles. The preferred squark and gluino
masses lie below 1000~GeV for $\tan \beta = 10$, with somewhat heavier
values for $\tan \beta = 50$, though still well within the reach of the
LHC~\cite{EHOW3}.

\section{Gravitino Dark Matter}

\begin{figure}[htb!]
\begin{center}  
\includegraphics[width=.48\textwidth]{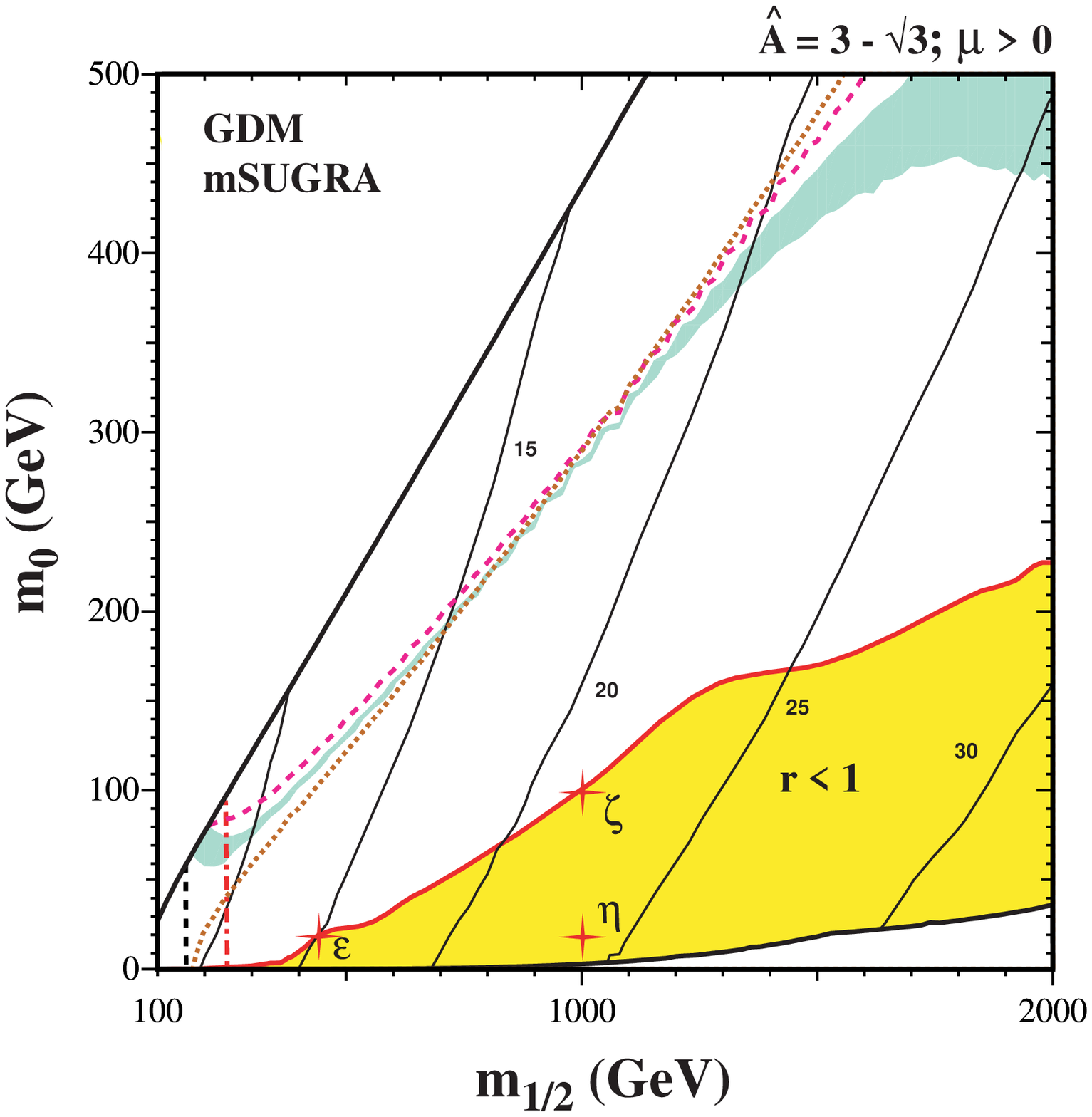}
\includegraphics[width=.48\textwidth]{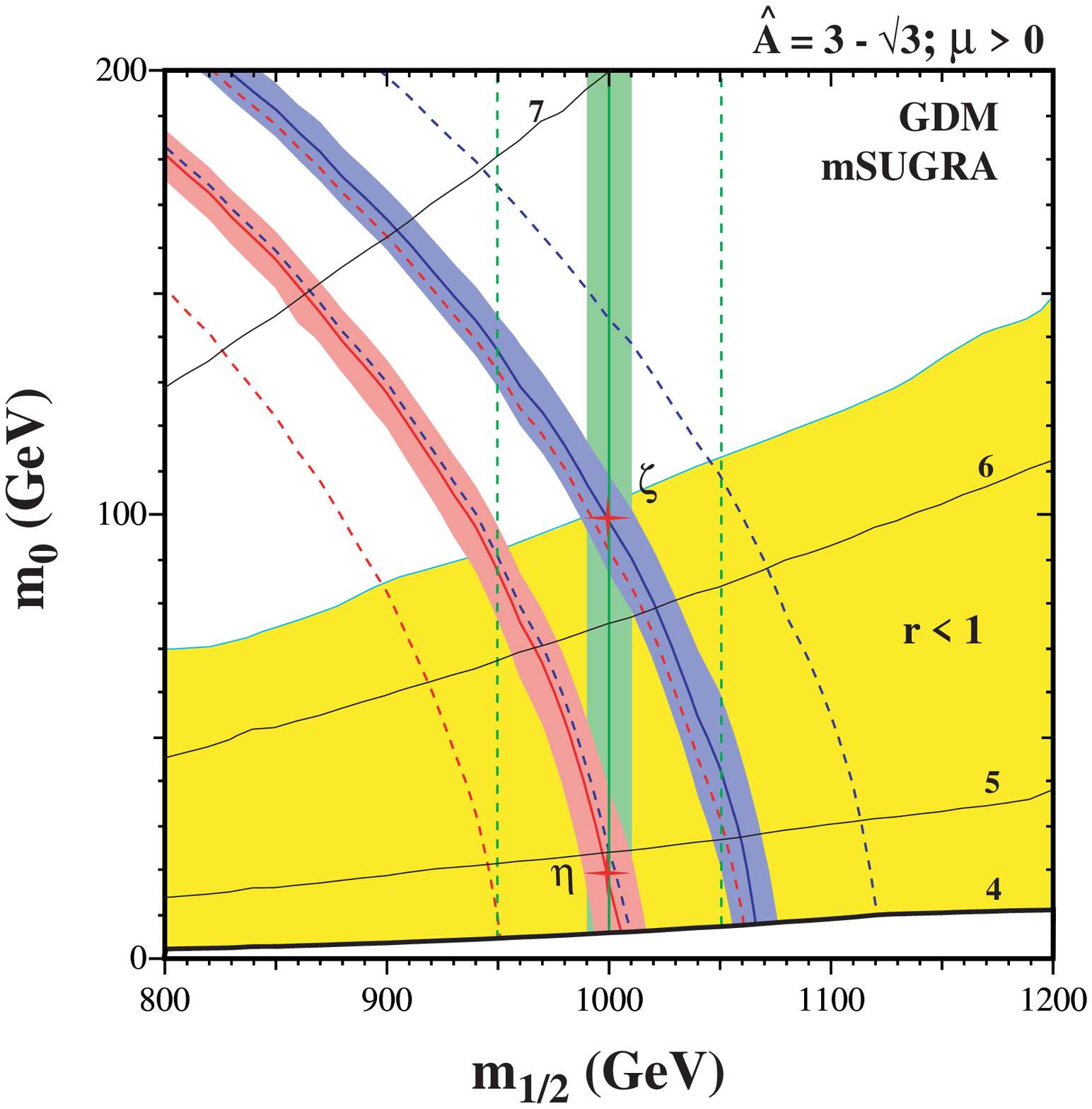}
\end{center}
\caption{
Left panel: The allowed region in the $(m_{1/2}, m_0)$ planes for minimal 
supergravity (mSUGRA) with a gravitino LSP [light (yellow) shaded
regions labelled $r < 1$], for $A \equiv {\hat A} m_0: {\hat A} = 3 -
\sqrt{3}$. The red crosses denote the benchmark GDM models 
$\epsilon, \zeta$ and $\eta$~\protect\cite{Bench3}. Right panel: The 
potential impacts on the determination of GDM parameters in the mSUGRA 
framework of prospective 
measurement errors of 1~\% and 5~\% for $m_{{\tilde \tau}_1}$
(diagonal bands and lines) and $m_{1/2}$ (vertical bands and lines),
shown as constraints in the $(m_{1/2}, m_0)$ plane~\protect\cite{Bench3}.
The smaller errors would enable the benchmark scenarios
$\zeta$ and $\eta$ to be distinguished, and the possible NLSP lifetime
to be estimated. The near-horizontal thin solid lines are labelled by the 
logarithm of the NLSP lifetime in seconds.
}
\label{fig:GDM}
\end{figure}

The above analysis assumed that the lightest supersymmetric particle (LSP)
is the lightest neutralino $\chi$, assuming implicitly that the gravitino
is sufficiently heavy and/or rare to have been neglected. This implicit
assumption may or may not be true in a minimal supergravity model, where
the gravitino mass $m_{3/2} = m_0$, as seen in
Fig.~\ref{fig:GDM}~\footnote{Minimal supergravity also relates the trilinear
and bilinear supersymmetry-breaking parameters: $A_0 = B_0 + 1$, thereby
fixing $\tan \beta$ as a function of $m_{1/2}, m_0$ and
$A_0$, see the contours in Fig.~\ref{fig:GDM}(b).}~\cite{GDM}. 
In this model, the gravitino mass is fixed throughout
the $(m_{1/2}, m_0)$ plane: there is a familiar WMAP strip where the
$\chi$ is the LSP, but there is also a wedge of parameter space where the
LSP is the gravitino. There is no way known to detect such astrophysical
gravitino dark matter (GDM), since the gravitino has very weak
interactions.

However, the LHC may have prospects for detecting GDM
indirectly~\cite{Feng,Nojiri,Bench3}.  In the GDM region, the lighter
stau, ${\tilde \tau}_1$, is expected to be the next-to-lightest sparticle
(NLSP), and may be metastable with a lifetime measurable in hours, days,
weeks, months or even years. The ${\tilde \tau}_1$ would be detectable in
CMS or ATLAS as a slow-moving charged particle. Staus that are
sufficiently slow-moving might be stopped in the detector itself, in some
external detection volume designed to observe and measure their late
decays into GDM~\cite{Feng,Nojiri}, or in the walls of the caverns
surrounding the detectors~\cite{Bench3}.

\section{The LHC and Ultra-High-Energy Cosmic Rays}

Historically, 
the two experiments with (until recently) the largest statistics for ultra-high-energy cosmic 
rays (UHECRs), 
AGASA~\cite{AGASA} and HiRes~\cite{HiRes}, have not agreed on their energy spectra above about $10^{19}$~eV and, 
specifically, whether there is a significant
number of events beyond the GZK cutoff due to interactions of primary UHECRs with the cosmic microwave background
radiation. The Auger experiment now has the second-largest statistics but 
does not yet have sufficient data to settle the issue~\cite{Auger}, 
though these should soon
be forthcoming. If there are super-GZK events, they might be due either to nearby astrophysical sources that have not
yet been identified, or (more speculatively) to the decays of metastable superheavy particles~\cite{topdown}. 
Normalizing the
energies of UHECRs requires understanding of the development of extensive air showers. At the moment, this is not
very well known, and models of shower development are not even able to tell us the composition of cosmic rays
with lower energies between $10^{15}$ and $10^{19}$~eV~\cite{UHECRs}.

The LHC is the accelerator that comes closest to reproducing the UHECR energy range, with a centre-of-mass energy
corresponding to $4 \times 10^{17}$~eV, in the range where the cosmic-ray composition is still uncertain. This
uncertainty would be reduced by better modelling of hadronic showers, which would in turn benefit from
measurements in the forward direction~\cite{UHECRs}.

Unfortunately, the LHC is currently not equipped to make good measurements in this kinematic region, where most
of the centre-of-mass energy is deposited. More instrumentation in the forward direction would be most welcome in
both CMS and ATLAS. This region is also of fundamental importance for our understanding of QCD, as I now explain.

\section{Back to Forward QCD}

We discussed earlier the success of perturbative QCD, and the accuracy
with which it could be used to calculate high-$p_T$ physics, thanks to the
structure functions provided by HERA data~\cite{PDF}, in particular.  The
simple parton description is expected, however, to break down at `small'
$x$ and `large' $Q^2$, due to saturation effects. At small $x$, there is a
large probability to emit an extra gluon $\sim \alpha_s {\rm ln}(1/x)$,
and the number of gluons grows in a limited transverse area. When the
transverse density becomes large, partons of size $1/Q$ may start to
overlap, and non-linear effects may appear, such as the annihilation of
low-$x$ partons. The Malthusian growth in the number of gluons seen at
HERA is eventually curbed by these annihilation effects when ${\rm
ln}(1/x)$ exceeds some critical $x$-dependent saturation value of $Q^2$.
At larger values of $x$, the parton evolution with $Q^2$ is
described by the usual DGLAP equations, and the evolution with ${\rm
ln}(1/x)$ is described by the BFKL equation. However, at lower values of
$x$ and large $Q^2$, a new description is need for the saturated configuration,
for which the most convincing proposal is the Colour-Glass Condensate
(CGC)~\cite{CGC}.

According to the CGC proposal, the proton wave function participating in
interactions at low $x$ and $Q^2$ is to be regarded as a classical colour
field that fluctuates more slowly than the collision time-scale. This
possibility may be probed in Gold-Gold collisions at RHIC and
proton-proton collisions at the LHC: the higher beam energy of LHC
compensates approximately for the higher initial parton density in
Gold-Gold collisions at RHIC. At central rapidities $y \sim 0$, effects of
the CGC are expected to appear only when the parton transverse momentum $<
1$~GeV. However, CGC effects are expected to appear at larger parton
transverse momenta in the forward direction when $y \sim 3$. Lead-Lead
collisions at the LHC should reveal even more important saturation
effects~\cite{KLN}.

\begin{figure}[htb!]
\begin{center}
\includegraphics[width=.48\textwidth]{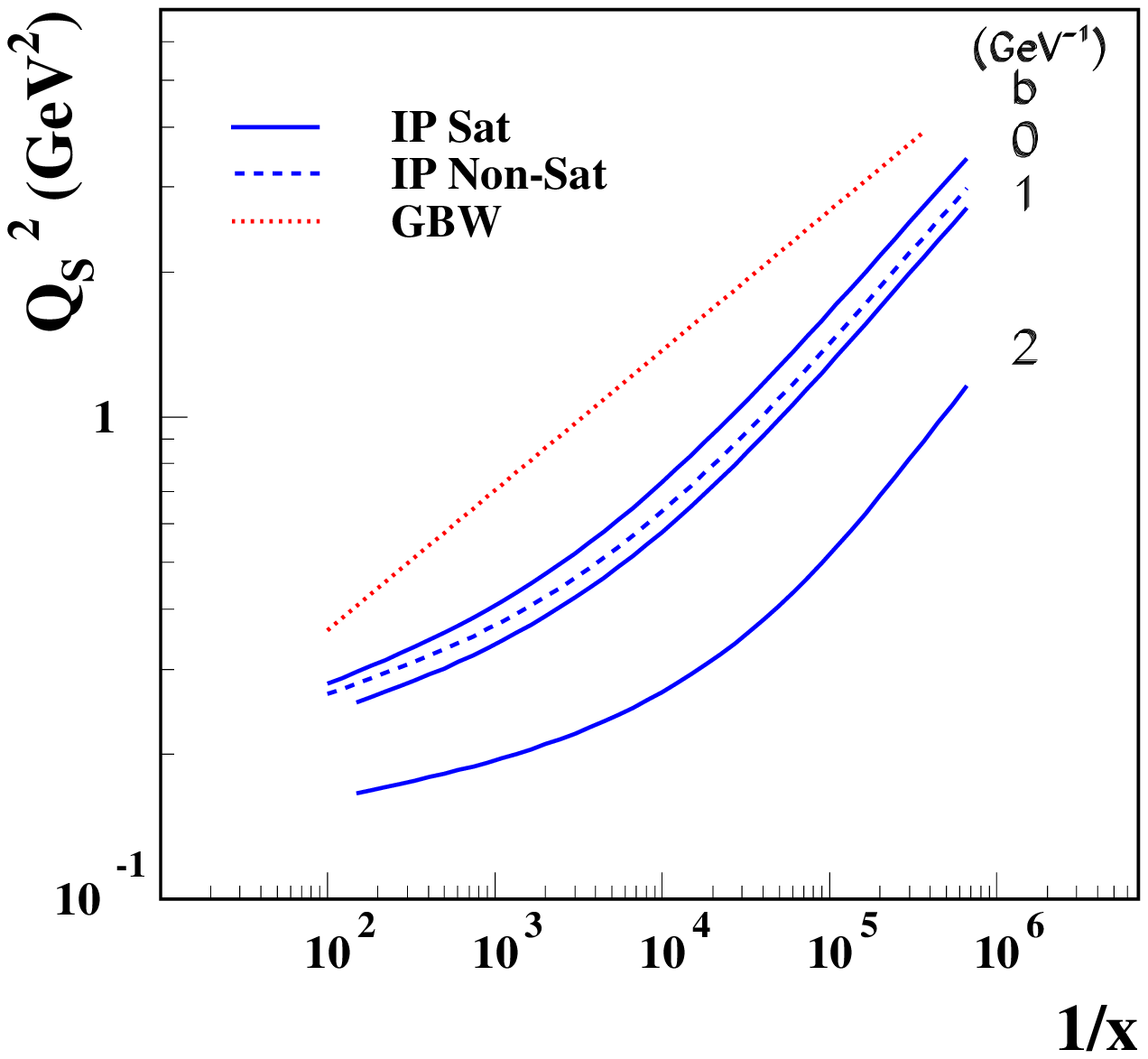}
\includegraphics[width=.48\textwidth]{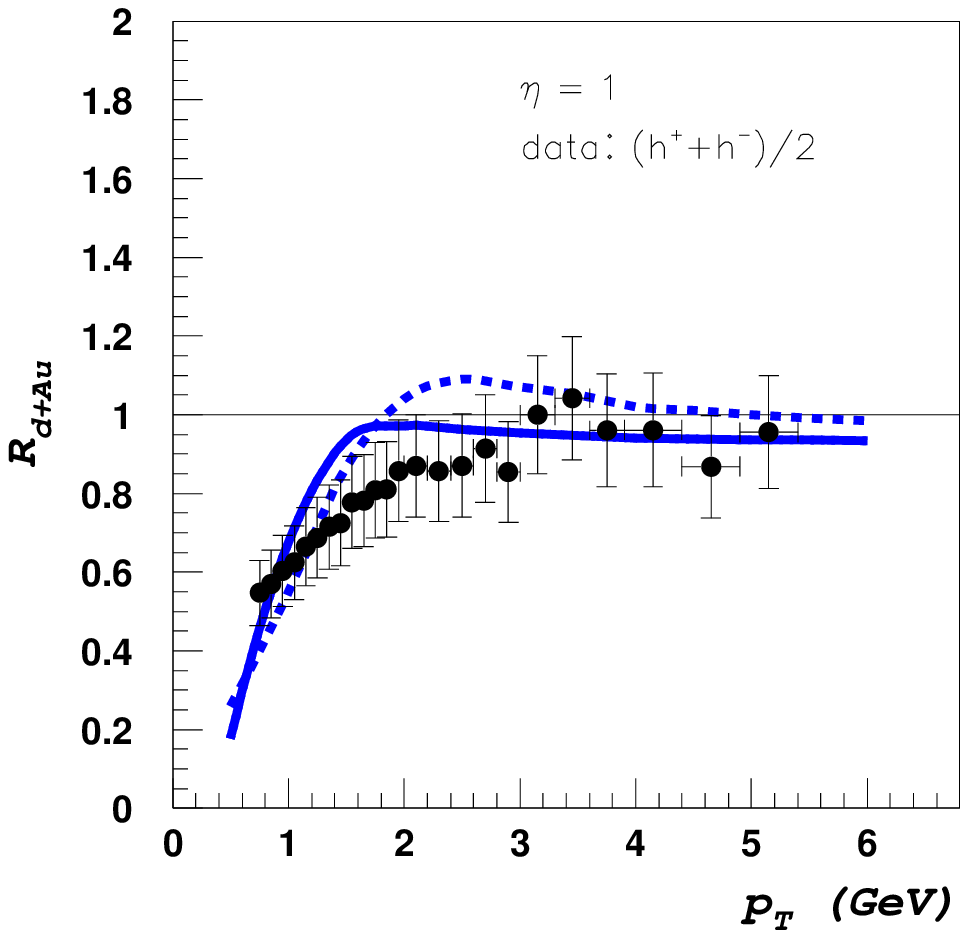}\\
\includegraphics[width=.48\textwidth]{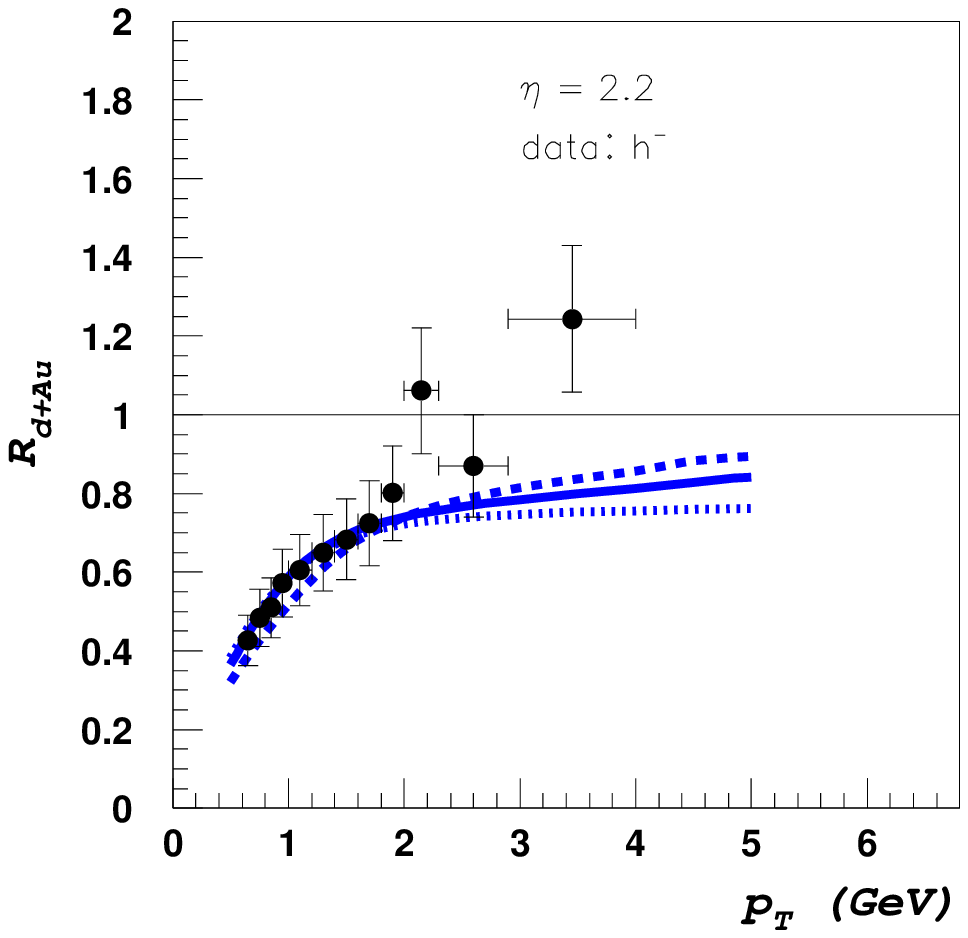}
\includegraphics[width=.48\textwidth]{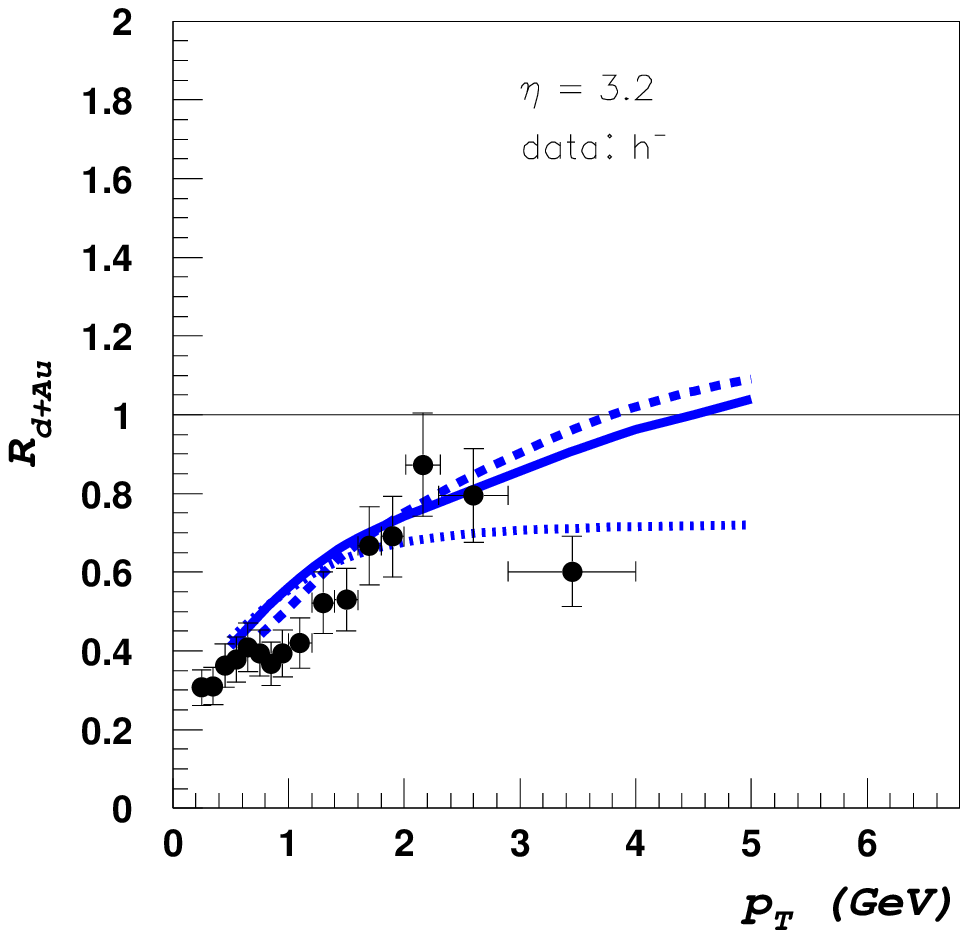}
\end{center}
\caption{
Top left panel: The parton saturation scale as a function of Bjorken $x$, 
extracted from HERA data in~\protect\cite{HERAsat}. Other three panels:
Nuclear modification factor $R_{dAu}$ of charged particles
for rapidities $\eta=1, 2.2, 3.2$~\protect\cite{BRAHMS}, compared with 
calculations from~\protect\cite{KKT}.
}
\label{fig:CGC}
\end{figure}

What is the experimental evidence for parton saturation? First evidence
came from HERA, and Fig.~\ref{fig:CGC}(a) displays an extraction of the
saturation scale from HERA data~\cite{HERAsat}. At RHIC, in proton-nucleus
collisions one expects the suppression of hard particles at large rapidity
and small angle compared to proton-proton collisions, whereas one expects
an enhancement at small rapidity, the nuclear `Cronin effect'. The
data~\cite{BRAHMS} from the BRAHMS collaboration at RHIC shown in
Fig.~\ref{fig:CGC}(b) are quite consistent with CGC
expectations~\cite{KKT}, but it remains to be seen whether this approach
can be made more quantitative than older nuclear shadowing ideas.

\section{New Physics in Diffraction?}

HERA has revealed a menagerie of different diffractive phenomena, opening
up a Pandora's box of possible new physics at the LHC. Classically one had
soft diffraction dissociation in peripheral proton-proton collisions, in
which one (or both) of the colliding protons would dissociate into a
low-mass system (or systems). HERA discovered an additional class of
diffractive events~\cite{HERAgap}, which may be interpreted~\cite{harddiff}
as a small colour dipole produced by an incoming virtual photon penetrates
the proton and produces a high-mass system. Additionally, one expects at
the LHC soft double diffraction, in which a peripheral proton-proton
collision produces a low-mass central system separated from each beam by a
large rapidity gap. Events with mixed hard and soft diffraction are also
possible at the LHC, as are events with multiple large rapidity gaps. The
LHC will certainly provide good prospects for deepening our understanding
of diffraction, building upon the insights being gained from HERA.

Double diffraction also offers the possibility of searching for new
physics in a relatively clean experimental environment containing, in
addition to Higgs boson or other new particle, only a couple of protons or
their low-mass diffraction-dissociation products~\footnote{New physics
might also be produced in other classes of diffractive events, but with
less distinctive signatures.}. The leading-order cross-section formula
is~\cite{Durham}:

$$
M^2 \; \frac{\partial^2 {\cal L}}{\partial y \partial M^2} = 4.0 \times 10^{-4}
\, \left [ \frac{\int^{l n \mu}_{lnQ_{min}} \; F_g (x_1, x_2, Q_T, \mu) dln Q_T}
{{\rm GeV}^-2} \right ]^2 \;
\left ( \frac{\hat{S}^2}{0.02} \right ) \;
\left ( \frac{4}{b {\rm GeV}^2} \right )^2 \;
\left ( \frac{R_g}{1.2} \right )^4
$$
where nominal values of the diffractive parameters are quoted in the
brackets. The gluon collision factor is currently inferred from HERA data
via different parameterizations of the integrated gluon distribution
function, and has an uncertainty of a factor of about two~\cite{Durham}.
Further analyses of HERA data, as well as future LHC data, would enable
the determination to be refined.

\begin{figure}[htb!]
\begin{center}
\includegraphics[width=.48\textwidth]{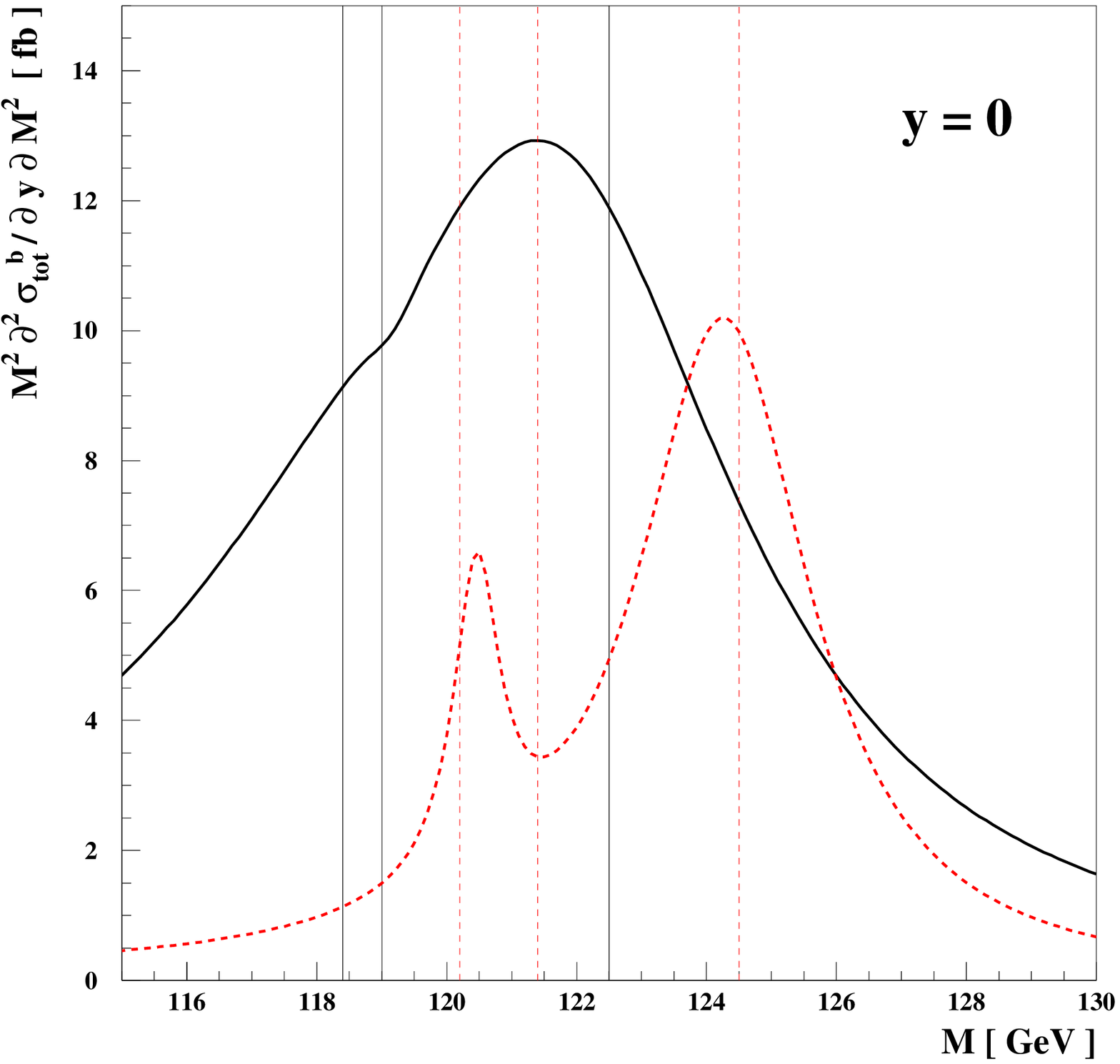}
\includegraphics[width=.48\textwidth]{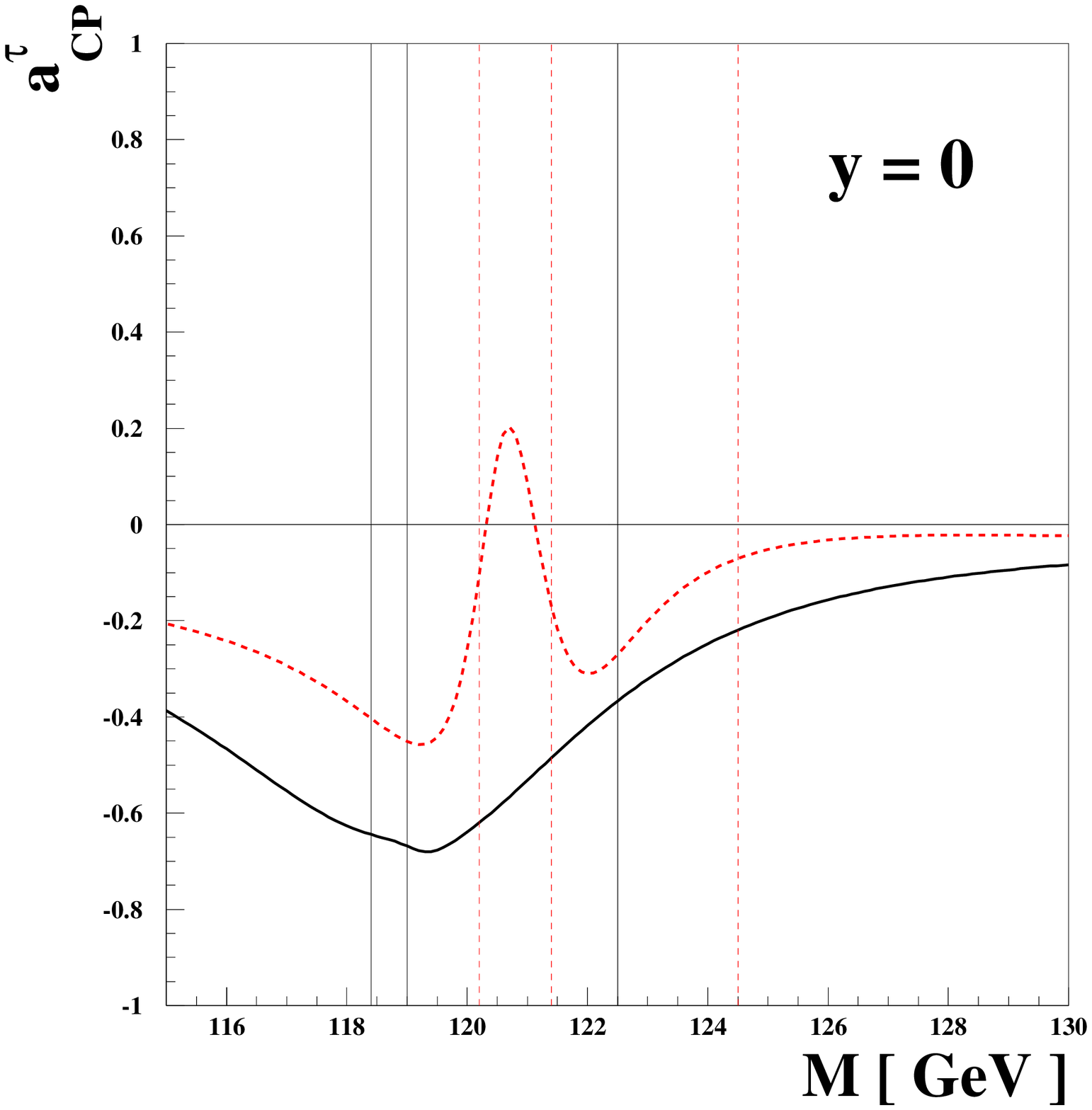}
\end{center}
\caption{
Left panel: The hadron-level cross section 
for the double-diffractive producion of Higgs  bosons decaying into $b$  
quarks. CP-violating three-way mixing  scenarios have been taken,
with the gluino phase $\Phi_3=-90^\circ$ (solid  lines) and 
$\Phi_3=-10^\circ$  (dotted line). The vertical lines  indicate the three 
Higgs-boson  pole-mass positions. Right panel: The CP-violating asymmetry 
$a^\tau_{\rm CP}$ observable in three-way mixing 
scenarios when Higgs bosons decay into $\tau$ leptons, using the same line 
styles~\protect\cite{ELP3}.
}
\label{fig:ELP3}
\end{figure}

The observation of diffractive Higgs production at the LHC would be a
challenge in the Standard Model, but the cross section is expected to be
considerably larger in the MSSM, particularly at large $\tan \beta$. One
of the enticing possibilities offered by supersymmetry is a set of novel
mechanisms for CP violation induced by phases in the soft
supersymmetry-breaking parameters~\cite{ELP}.  These would show up in the
MSSM Higgs sector, generating three-way mixing among the neutral MSSM
Higgs bosons. This might be observable in inclusive Higgs production at
the LHC~\cite{ELP}, but could be far more dramatic in double diffraction.
Fig.~\ref{fig:ELP3}(a) displays the mass spectrum expected in double
diffraction in one particular three-way mixing scenario~\cite{ELP3}:  it
may exhibit one or more peaks that do not coincide with the Higgs masses.
Analogous structures may also be seen in CP-violating asymmetries in $H_i
\to \tau^+ \tau^-$ decay, as seen in Fig.~\ref{fig:ELP3}(b). These
structures could not be resolved in conventional inclusive Higgs
production at the LHC, but may be distinguished in exclusive double
diffraction by exploiting the excellent missing-mass resolution $\sim
2$~GeV that could be provided by suitable forward
spectrometers~\cite{FP420}.

\section{Summary}

We do not know what the LHC will find - maybe there will be no supersymmetry and we will observe mini-black-hole
production instead! However, whatever the physics scenario, HERA physics will provide crucial inputs, for example
via measuring the parton distributions that will be crucial for searches for new physics such as the Higgs boson,
or via the observation of saturation effects that will be important for forward physics, or via measurements of
diffraction.

Forward physics is a potentially exciting area of LHC physics that is not covered by the present detectors. HERA
and RHIC suggest that parton saturation and the Colour Glass Condensate may be observable here, understanding of
forward physics is essential for the modelling of cosmic-ray air showers and hence determining the spectrum of
ultra-high-energy cosmic rays, and diffractive events related to those observed by HERA may be a valuable tool
for discovering new physics such as Higgs production. There is still plenty of room at the LHC for novel
experimental contributions~\cite{FP420}.

\end{document}